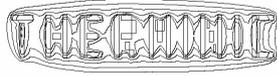

# THE METHOD OF NON-LINEAR DISTORTIONS ELIMINATION IN PHOTOACOUSTIC INVESTIGATION OF LAYERED SEMICONDUCTOR STRUCTURE

*Zbigniew Suszyński [1], Radosław Duer [2], Mateusz Kosikowski [3]*

University of Technology Koszalin, Department of Electronics and Computer Science,
Śniadeckich 2, 75-453 Koszalin, Poland,

email: [1] zas@man.koszalin.pl , [2] rduer@man.koszalin.pl, [3] kazuchiro@interia.pl

## ABSTRACT

The one of the most current issues in semiconductor technology is a problem of assessment of the quality of adhesion between layers in produced semiconductor structure. Weak adhesion highly affects thermal impedance of the interface area and it conditions the heat abstraction from the structure. An excessive increase of temperature during operation of device leads to its damage so at this point of view the quality of adhesion is one of the most important parameters of the reliability of semiconductor devices.

## 1. INTRODUCTION

Magnificent tools for the quality of adhesion control are thermal wave methods. Among them, the photo-acoustic (PA) technique in broad-band modulation mode can be distinguished. For this technique, The photo-acoustic phenomenon is being utilitized within. The modulated laser beam is used to excite a variable heat flux inside the structure so it changes the pressure of constant volume of gas closed inside photo-acoustic chamber. The sound generated as a result of pressure changes is being registered with a microphone. This technique allows measuring the temperature response fast and with relatively low costs [1-4].

However, the PA technique, as any other indirect method of temperature measurement is characterized by high level of linear distortions [5]. Thus the correct analysis of thermal properties basing upon the photo-acoustic measurements is possible under two conditions: the reduction of linear distortion influence and elimination of non-linear distortions.

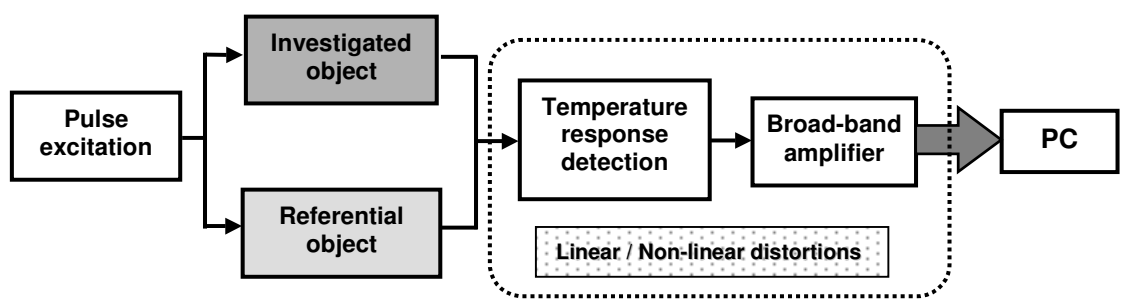

*Fig.1. General scheme of photo-acosutic technique in broad-band excitations mode.*

The first condition can be assured using a comparative method of analysis, for example the contrast method. Regarding the second condition, consideration of the presence and the influence of non-linear distortions as well as the elimination of that are presented in this contribution.

## 2. THE OBJECT OF INVESTIGATION AND THE REFERENCE OBJECT

High power thyristor structure was used as an object of investigation and a sample of passivated glass was assumed as a reference object. The investigated object consisted of 3 layers: silicon – eutectics – molybdenum.

The thickness and thermophysical properties of each layer are presented in Tab. 1.

In the case of weak adhesion between layers, an additional layer of high thermal impedance and strongly diversified thermal parameters can be formed. It was assumed that it is a layer of air.

## 3. THE EXPERIMENT

The experiment was performed using the photo-acoustic measurement set-up presented in Fig. 3. Photo-acoustic (PA) images were acquired in pulse excitations mode. This mode allows significantly reduce the time of experiment because the spectra of the temperature response contain in this case complete information about thermal properties of object. Laser diode was used as a source of excitation. The intensity of a laser beam was modulated by a rectangular pulse (duration $t_0 = 0.1$s, period $T = 2.5$s). PA signal was registered synchronously with switching the laser on with time resolution of 1ms. The registration of PA signal was performed for both the investigated and the reference objects. Fig.4 presents the PA signal registered during the experiment. In order to investigate the presence of non-linear distortions as well as their possible influence on PA signal, PA responses were registered for different values of laser diode current ranging from 0.86A to 1.26A. The signal, registered for the excitation with 1.26A laser diode current, is highly distorted. In this regime the amplifier was saturated, however it was necessary to investigate the signal in this range either because it was assumed that non-linear distortions should be seen in this regime for sure. In effect spectra of all signals are slightly different and it will be evidenced farther.

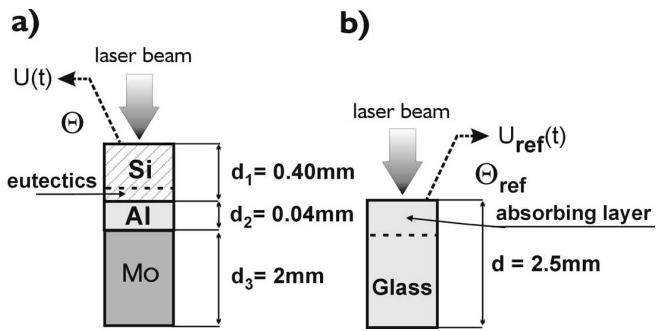

*Fig. 2. A cross-section of the investigated structure and the reference object*

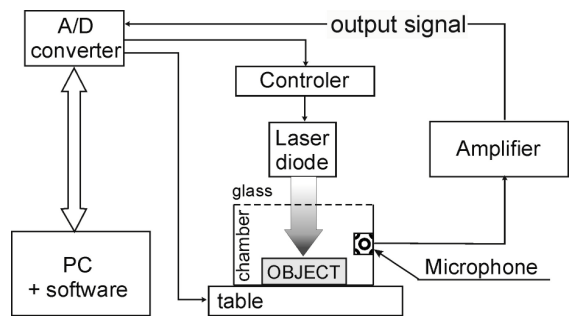

*Fig. 3. PA setup used for the experiment*

| Number of layer | Thickness of layer [mm] | Material | Density of material $\rho$ | Thermal conductivity coef. $\lambda$ | Specific heat c |
|---|---|---|---|---|---|
| d1 | 0,40 | siliconium | 2330 | 140 | 678 |
| d2 | 0,04 | aluminium | 2700 | 237 | 902 |
| d3 | 2 | molibdenum | 10250 | 138 | 250,2 |
| d0 | $10^{-8} \div 10^{-5}$ | air | 1,205 | 0,0259 | 1005 |

*Tab.1. The thickness and thermophysical properties of each layer of investigated structure.*

## 4. THE ANALYSIS OF EXPERIMENTAL RESULTS

As it was mentioned before, the successful elimination of the influence of PA signal linear distortion is the crucial condition of the analysis of thermal properties of investigated structure. One of the most effective methods for that is a use of contrasts characteristics in frequency domain [6]. The contrast's definitions presented (1, 2) were evaluated basing upon the spectra of registered PA signal. In order of that it is necessary to determine the spectrum of photo-acoustic signal using DFT transformation [7].

The comparison of the spectrum characteristic of the PA signal for different values of supply current is presented in Fig.5. The changes of supply power results in the relocation of low frequency harmonic energy into the range of high frequencies. The changes of the spectra caused by non-linear distortions seem to be not so big but they may cause the substantial differences in the contrasts characteristics. At the same time, it can be observed that the spectra are unstable in time.

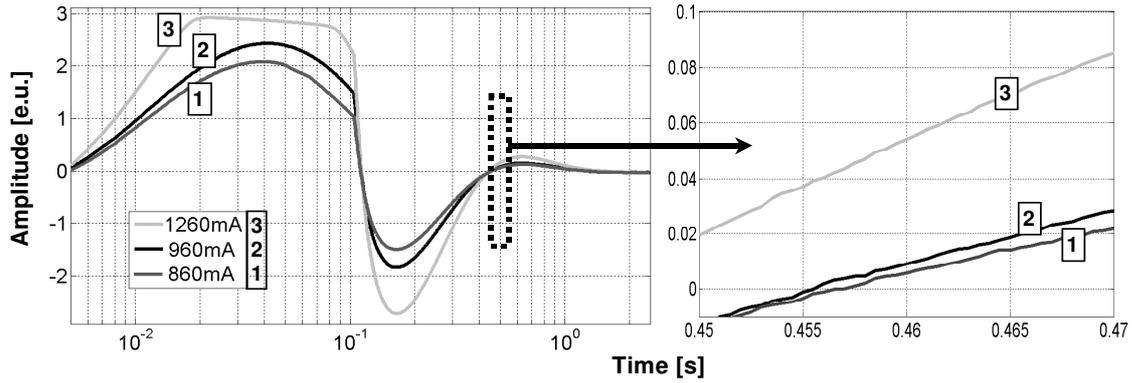

*Fig. 4. The photoacoustic signal registered for the reference object.*

$$AC(\omega) = \frac{|U(\omega)| - |U_{REF}(\omega)|}{|U_{REF}(\omega)|} = \frac{|\Theta(\omega) \cdot \dot{K}| - |\Theta_{REF}(\omega) \cdot \dot{K}|}{|\Theta_{REF}(\omega) \cdot \dot{K}|} = \frac{|\Theta(\omega)| - |\Theta_{REF}(\omega)|}{|\Theta_{REF}(\omega)|} \cdot 100\% \qquad (1)$$

$$PhC(\omega) = -\{\arg(U(\omega)) - \arg(U_{REF}(\omega))\} = -(\varphi + \Delta\varphi - \varphi_{REF} - \Delta\varphi) = -(\varphi - \varphi_{REF}) \qquad (2)$$

**where:**

$\dot{K}$ is the complex transmittance of measurement setup; $U$ and $U_{REF}$ are output signals for the investigated and reference areas respectively; $|\Theta|$, $|\Theta_{REF}|$, $\varphi$ and $\varphi_{REF}$ are amplitudes and phases of temperature disturbance for both areas, $\Delta\varphi$ is phase lag introduced by measurement setup.

A) amplitude spectrum

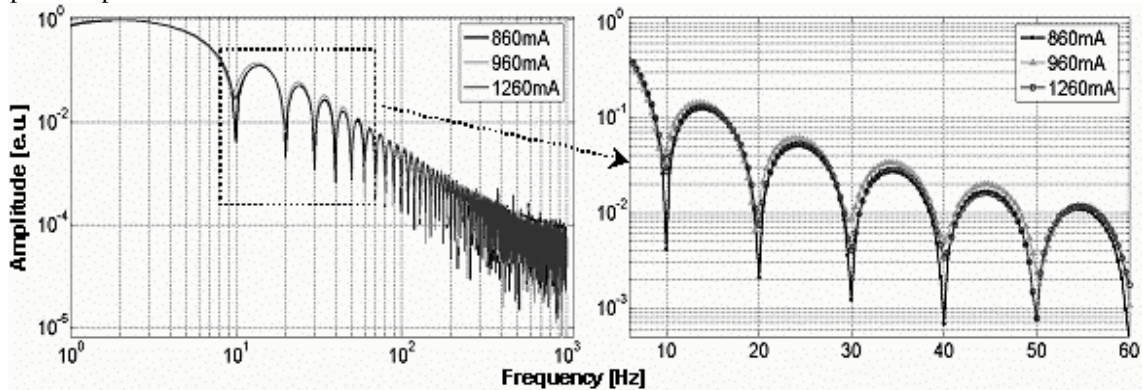

B) phase spectrum

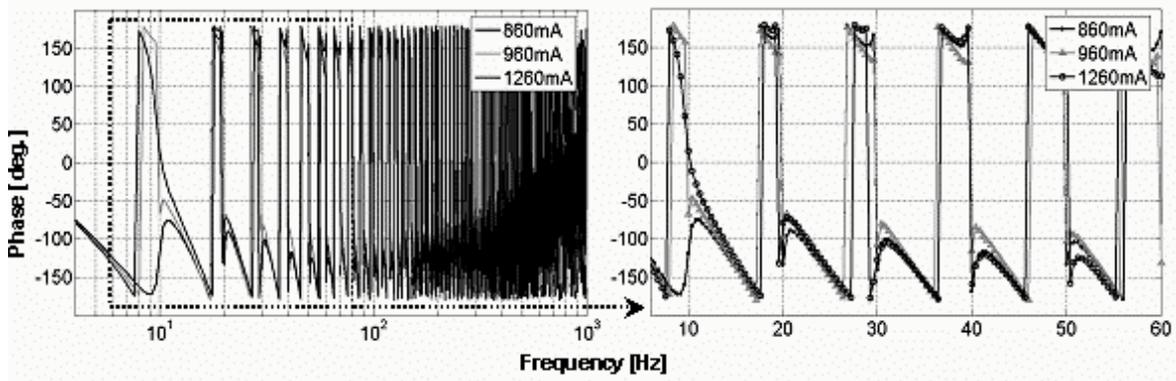

*Fig.5. Frequency spectra of registered photo-acoustic signal.*

It seems to be caused by a relation between the level of non-linear distortions and low frequency acoustic disturbances registered by the microphone. Relatively high amplitude of low frequency disturbance influences on the range of all photo-acoustic signal changes and for this its spectrum is unstable with time.

Fig. 6 presents the amplitude and phase contrasts characteristics obtained from (1) and (2). Both of contrast characteristics are strongly distorted especially for frequencies being multiple of 10Hz. It is caused by near to zero value of spectra for multiple of 10Hz.

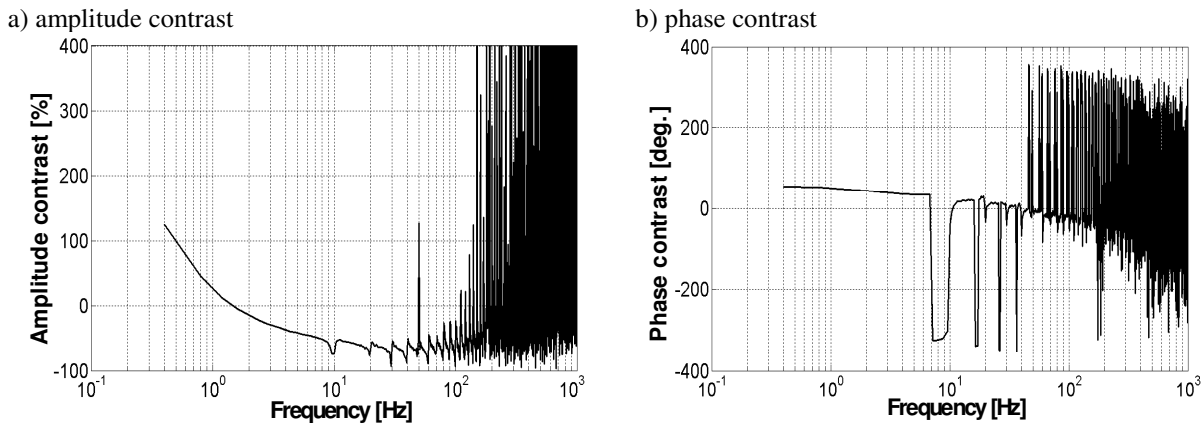

*Fig.6. Contrasts characteristics evaluated basing upon the photoacoustic signal*

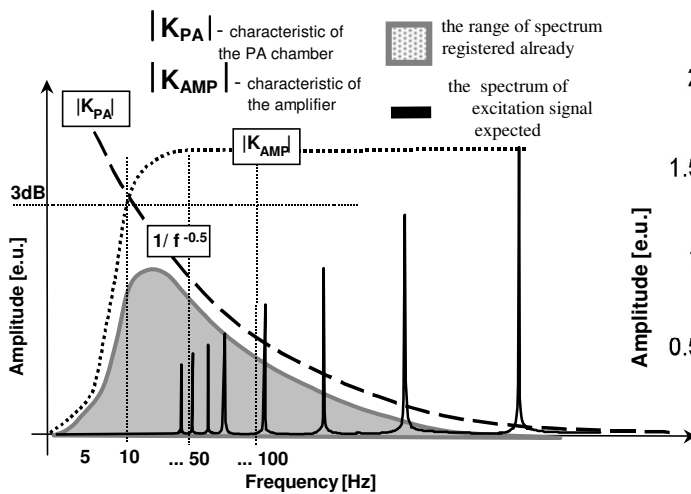

*Fig.7. The spectrum of desired broad-band excitation signal*

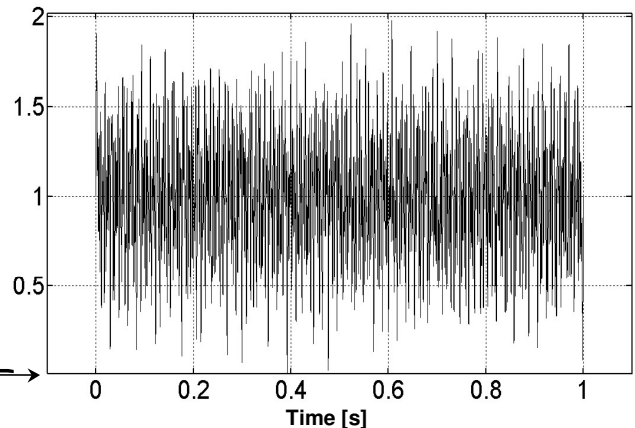

*Fig.8. The broad-band excitation signal in time.*

This value is in the denominator of eq.(1) and for this a small fluctuation of its can result in a huge changes of amplitude contrast. This phenomenon is all the more significant that the signal-to-noise ratio nearby these frequencies is essentially smaller then for other harmonics. The level of disturbances is growing with frequency. It comes from the fact that the PA signal decreases with the growing frequency as 1/f. It means that signal-to-noise ratio decreases with frequency too. Simultaneously, the amplitude and the phase contrast characteristics change with power of excitation, which is caused by non-linear distortions of PA signal.

## 5. CONCLUSIONS

The results presented above lead to the conclusion that the analysis of thermal properties of the investigated structure is possible under two conditions:

1. Increase of the noise-to-signal ratio;
2. The elimination of non-linear distortions of the pa signal by performing the measurements in the range of exciation's power for which the non-linear distortions are not present.

In authors opinion, fulfilling these two opposite requirements is possible only with the use of special kind of excitation signal. Such a signal should be characterised by discrete and finite spectrum consisted of selected frequencies. Simultaneously, the amplitude of each harmonic should be distributed aversely to the frequency characteristics of components of the measurement setup. That way, the energy of excitation will be congregated in specific frequencies and the amplitude of signal will be distributed more efficiently.

An example of the spectrum of desired excitation signal is presented in Fig.7. Such modification of photoacoustic technique is cheap and seems to be effective. However it is necessary to point that the successful application of the broad-band excitation signal proposed is conditioned by the performance of the light source.